\documentclass[apjl]{emulateapj}
\usepackage{amssymb}
\usepackage{graphicx,graphics}% Include figure files
\usepackage{epsfig}% Include figure files
\usepackage{dcolumn}% Align table columns on decimal point
\usepackage{bm}% bold math
\usepackage{natbib}
\usepackage{times}
\usepackage{ulem}
\def\kmm#1  {{\bf [KMM:~ #1]~}}
\def\new#1 {{\bf #1 }}
\def\cut#1 {\sout{#1} }
\newcommand{\htwo}{\ensuremath{\rm H_2}}
\def\mgtwo{Mg{\sc ii}}
\newcommand{\cm}{cm$^{-2}$}
\newcommand{\hii}{H{\sc i}~21cm}

\newcommand{\beq}{\begin{equation}}
\newcommand{\eeq}{\end{equation}}

\newcommand{\hi}{H{\sc i}}
\newcommand{\kms}{km~s$^{-1}$}

\shorttitle{A blind GBT survey for molecular absorption}
\shortauthors{Kanekar et al.}
\begin{document}
\title{A blind Green Bank Telescope millimetre-wave survey for redshifted molecular absorption}

\author{N. Kanekar\altaffilmark{1},
A. Gupta\altaffilmark{1,2},
C. L. Carilli\altaffilmark{3},
J. T. Stocke\altaffilmark{4},
K. W. Willett\altaffilmark{5}
}
\altaffiltext{1}{National Centre for Radio Astrophysics, 
TIFR, Ganeshkhind, Pune - 411007, India; nkanekar@ncra.tifr.res.in}
\altaffiltext{2}{Indian Institute of Science Education and Research, Mohali, India}
\altaffiltext{3}{National Radio Astronomy Observatory, 1003 Lopezville Road, Socorro, NM87801, USA}
\altaffiltext{4}{CASA, Department of Astrophysical and Planetary Sciences, University of Colorado, 389-UCB, Boulder, CO 80309, USA}
\altaffiltext{5}{School of Physics \& Astronomy, University of Minnesota, 116 Church St. SE,
Minneapolis, MN 55455, USA}

\begin{abstract}
We present the methodology for ``blind'' millimetre-wave surveys for redshifted molecular 
absorption in the CO/HCO$^+$ rotational lines. The frequency range $30-50$~GHz appears 
optimal for such surveys, providing sensitivity to absorbers at $z \gtrsim 0.85$. It is 
critical that the survey is ``blind'', i.e. based on a radio-selected sample, including
sources without known redshifts. We also report results from the first large survey 
of this kind, using the Q-band receiver on the Green Bank Telescope (GBT) to search for 
molecular absorption towards 36~sources, 3 without known redshifts, over the frequency 
range $39.6 - 49.5$~GHz. The GBT survey has a total redshift path of $\Delta z \approx 24$, 
mostly at $0.81 < z < 1.91$, and a sensitivity sufficient to detect equivalent 
${\rm H_2}$ column densities $\gtrsim 3 \times 10^{21}$~cm$^{-2}$ in absorption at 
$5\sigma$ significance (using CO-to-${\rm H_2}$ and HCO$^+$-to-${\rm H_2}$ conversion 
factors of the Milky Way). The survey yielded no confirmed detections of molecular 
absorption, yielding the $2\sigma$ upper limit $n(z=1.2) < 0.15$ on the redshift number density 
of molecular gas at column densities $N({\rm H_2}) \gtrsim 3 \times 10^{21}$~cm$^{-2}$.

\end{abstract}


\maketitle
\section{Introduction} 
\label{sec:intro}

Molecular gas is an important constituent of the interstellar medium, playing a critical 
role in galaxy evolution via its influence on star formation. However,
little is known about the cosmological evolution of molecular gas in galaxies. This is 
primarily due to the difficulty in detecting molecular hydrogen, $\htwo$, the dominant 
molecular species. Tracers of $\htwo$ like CO, HCO$^+$ and OH \citep[e.g.][]{liszt96,kanekar02,burgh07}
hence offer the best means to first detect, and then study evolution in, the molecular phase.

Emission studies of molecular gas in redshifted galaxies have mostly been carried out using 
CO transitions. Such studies have provided interesting information on massive objects, 
the sub-mm galaxies, B$z$K galaxies, and high-$z$ quasars \citep[e.g.][]{carilli13}. 
However, the fact that emission line strengths fall like the inverse square of the 
luminosity distance has meant that it has so far been difficult to detect CO emission 
in ``normal'' galaxies \citep[e.g.][]{wagg09,wagg12a}, except for highly lensed systems 
\citep[e.g.][]{baker04,riechers10}.

In contrast to emission surveys, the sensitivity of surveys for molecular absorption 
against active galactic nuclei does not decrease with redshift. Absorption-selected galaxies 
are thus more likely to be representative of the normal galaxy population. The bulk of 
the molecular phase in typical galaxies at any redshift is also likely to be excitationally 
cold and easier to detect in absorption. Molecular absorption studies can also provide 
important probes of cosmological evolution. The wealth of radio rotational lines allows 
detailed characterization of physical and chemical conditions in the absorbing gas 
\citep[e.g.][]{henkel05,henkel08,bottinelli09}. The relative strengths of different 
absorption transitions of species like H$_3$CN, where the excitation is dominated by 
the cosmic microwave background (CMB), can be used to determine the CMB temperature 
\citep[e.g.][]{henkel09,muller13}. Comparisons between the redshifts of different 
molecular (and atomic) transitions in an absorber can be used to test for cosmological 
evolution in the fundamental constants of physics 
\citep[e.g.][]{drinkwater98,kanekar10b,kanekar11,kanekar12,bagdonaite13}.

Searches for molecular absorption at high redshifts, $z > 1$, have so far been mostly
carried out via optical absorption spectroscopy, targetting the redshifted ultraviolet 
${\rm H_2}$ lines in damped Lyman-$\alpha$ systems \citep[DLAs; e.g. ][]{wolfe05}. Such studies 
have obtained detection fractions of $\approx 15$\% for DLAs at $z > 1.8$, with very 
low derived ${\rm H_2}$ fractions, 
$f \equiv 2N({\rm H_2})/[2N({\rm H_2}) + N({\rm HI})] \sim 10^{-6} - 0.1$
\citep[e.g.][]{ledoux03,noterdaeme08}, with typical limits of $f \lesssim 10^{-5}$. 
These results are not very surprising as a high molecular hydrogen column density would 
imply the presence of large amounts of dust, which would obscure the background quasar 
at ultraviolet and optical wavelengths, making it very hard to even determine the 
quasar redshift, let alone detect absorption features in the spectrum.

Surveys for molecular absorption must hence be carried out at radio wavelengths, 
to ensure no bias against dusty sightlines. An unbiased radio absorption survey would 
provide a census of molecular absorbers as a function of redshift, allowing one to study 
the evolution of the cosmological mass density in molecular gas, as has been possible for 
atomic gas through optical surveys for DLAs \citep[e.g.][]{wolfe86}. Unfortunately, the 
narrow fractional bandwidth hitherto accessible to radio spectrometers has made it difficult 
to carry out such surveys for molecular absorption. As a result, only five radio molecular 
absorbers are currently known at cosmological distances, all at $z < 0.9$ 
\citep{wiklind95,wiklind96,wiklind96b,wiklind97,kanekar05}. Four of these were detected in 
targeted searches, usually based on the redshift of \hii\ absorption 
\citep[e.g.][]{carilli92,carilli93,kanekar03d}. The $z \sim 0.886$ towards PKS1830$-$21 
remains the only molecular absorber found via a blind search \citep[][]{wiklind96b}.
In this paper, we report the first large ``blind'' survey for redshifted molecular absorption, 
using the wide-bandwidth capabilities of the AutoCorrelation Spectrometer (ACS) on the 
Green Bank Telescope (GBT).

%Such studies have provided the most accurate measurement of the CMB temperature at 
%any redshift, $T_{\rm CMB} = 5.08 \pm 0.10$~K from the $z = 0.886$ gravitational lens 
%towards 1830$-$211 \citep{muller13}.  
%Such studies have yielded amongst the most stringent constraints on changes in the 
%proton-electron mass ratio and the fine structure constant 
%\citep[e.g.][]{kanekar10b,kanekar11,kanekar12,bagdonaite13}.

\section{A radio survey for molecular absorption}
\label{sec:survey}

The redshift sensitivity of an absorption survey is quantified by the redshift path density
$g(z)$, defined so that $g(z)dz$ gives the number of sightlines for which optical depths 
above the threshold sensitivity would have been detected in the interval
$z$ to $z+dz$ \citep[e.g.][]{lanzetta91}. The total survey redshift path is given 
by $\Delta z = \int_0^\infty g(z) dz$, essentially the sum over the redshift ranges 
for each target over which the survey is capable of detecting absorption. The primary 
goal in an absorption survey is to maximize $\Delta z$ for detectable absorption, at a 
given sensitivity to column density or optical depth \citep[e.g.][]{wolfe86}. 

A radio molecular absorption survey is best done using the strong rotational lines of CO 
(at $\approx 115$, $230$, $345$,...~GHz), and HCO$^+$ (at $\approx 89$, $178$,
$267$,...GHz). While CO is far more abundant (by about a factor of 1000) than HCO$^+$ in the 
interstellar medium, the HCO$^+$ rotational lines have a higher Einstein-A coefficient than 
the corresponding CO rotational lines, implying that the HCO$^+$ optical depth along a sightline 
is typically about an order of magnitude lower than the CO optical depth. The efficiency of a 
mm-wave molecular absorption survey is thus significantly increased because observations covering
a given frequency range are sensitive to absorption in multiple redshift ranges, corresponding 
to the different CO and HCO$^+$ transitions. For example, an observing frequency 
of 40~GHz corresponds to redshifts $z \approx 1.22$~(HCO$^+ 1-0$), $\approx 1.88$~(CO~$1-0$), 
$\approx 3.46$~(HCO$^+ 2-1$), $\approx 4.76$~(CO~$2-1$), etc. One can thus choose the observing 
frequency band so as to maximize the redshift range towards a background source, thus increasing
the survey efficiency. Taking into account the frequency dependence of 
the atmospheric opacity, the best observing frequency range for a radio molecular absorption 
survey appears to be $\approx 31-49$~GHz. This would be sensitive to absorption in the 
redshift ranges $0.82 < z < 1.88$ (HCO$^+ 1-0$), $1.35 < z < 2.72$ (CO~$1-0$), 
$2.64 < z < 4.75$ (HCO$^+ 2-1$), $3.70 < z < 6.44$ (CO~$2-1$), etc. In other words, a 
survey covering $31-49$~GHz would be able to detect absorption by galaxies at {\it all 
redshifts $z \gtrsim 0.82$} in the CO and HCO$^+$ lines. Further, the relatively small frequency 
range implies fewer frequency settings per source and thus a high survey efficiency. 
We hence originally aimed to use the GBT Ka- and Q-band receivers and the ACS to search 
for redshifted absorption in the frequency range $31-49$~GHz.

In passing, we note that a similar absorption survey is possible at higher frequencies, in 
the 2-mm or 3-mm bands \citep[e.g.][]{wiklind96b}. For example, the observing frequency 
range $70-110$~GHz provides access to {\it all redshifts $z > 0$} in the CO and HCO$^+$ 
lines. However, a higher observing frequency implies both a larger frequency range 
(i.e. more frequency settings) to cover the same redshift path and weaker background 
sources. Further, using a high observing frequency would effectively target the high-J
rotational lines for higher redshift absorbers. In normal galaxies, the lower CO/HCO$^+$ 
rotational levels are likely to be highly populated \citep[e.g.][]{fixsen99}, implying that 
the low-J rotational transitions should yield the strongest absorption lines. 
All of these indicate that the frequency range $31-49$~GHz is likely to be the best 
for molecular absorption surveys, despite the fact that it only allows access to relatively
high redshifts, $z \gtrsim 0.8$. One might extend the survey coverage to lower redshifts 
by additional observations in the 3-mm band.

\section{The target sample}
\label{ref:sample}

For a molecular absorption survey, it is critical that the target selection be based 
on radio criteria. This is because the sample should contain no bias against optically 
faint active galactic nuclei (AGNs), which might be dim due to extinction by dust 
associated with a molecular absorber along the sightline. It would also be useful if 
the sample were selected from a radio survey as close to the observing frequency 
($\approx 40$~GHz) as possible, to reduce uncertainties in the source flux density. 

Unfortunately, at the time of the GBT observations (2006), there were no deep all-sky 
radio surveys at high frequencies, $\approx 40$~GHz, in the literature.  We hence chose 
to construct our main target sample from the well known 1~Jy sample \citep[][]{kuhr81}. 
This covers the whole sky excluding the Galactic plane (i.e. $b > 10^\circ$) and the 
Magellanic clouds and is expected to be complete to a flux density of $1$~Jy at 
5~GHz. We restricted our targets to declinations $\geq -30^\circ$ so that all sources 
could be observed at a reasonable elevation at the GBT. Finally, we used redshift 
information from the literature to exclude all sources with redshifts $z < 1.2$, 
to ensure sufficient redshift path ($\Delta z_{min} \sim 0.35$) for even the lowest 
redshift targets. We emphasize that sources without known redshifts were retained 
in the sample. The survey is hence not biased against targets possibly obscured by 
dust and, hence, without optical redshifts. Finally, while the statistical analysis 
will be restricted to the 1~Jy sample, we also augmented the sample with six sources 
without known redshifts from the UCSD survey \citep{jorgenson06}. Our initial sample 
consisted of 113~targets, 93 with measured redshifts between $z = 1.2$ and $z = 3.522$ 
and 20 without known redshifts.

\section{Observations, data analysis and results}
\label{sec:data}

\begin{figure}
\centering
\includegraphics[scale=0.4]{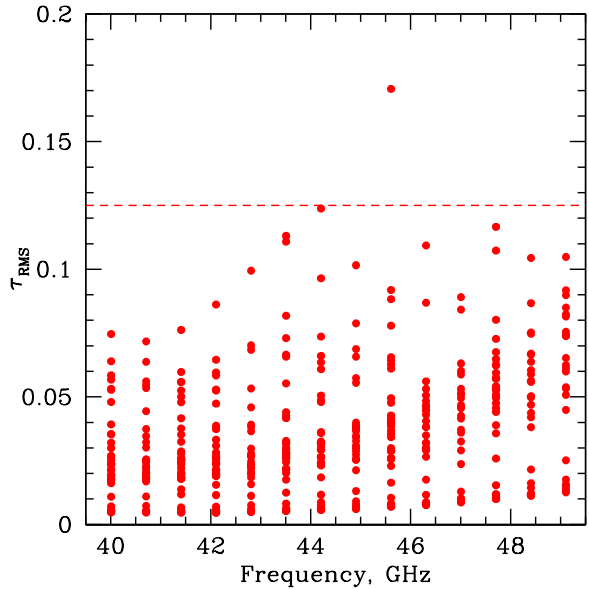}
\caption{The distribution of RMS optical depth noise as a function of frequency. 
Each point corresponds to a single 800~MHz spectrum; there are 438 such spectra 
towards 36~sources.}
\label{fig:rms}
\end{figure}

Before embarking on the GBT spectroscopy, we used the Very Large Array (VLA) in 
its D-configuration (proposal AK619) in 2005 to obtain 43~GHz flux densities for 
86~sources of the sample. The VLA observations used 1-minute on-source integrations 
and two 50-MHz IF bands, with the data analysed in {\sc aips}. The flux densities of 
the remaining 27 sources were obtained from the literature. These flux densities were 
used to plan the GBT spectroscopy, with observing times chosen so as to achieve 
relatively uniform optical depth sensitivity in the CO~$1-0$ and HCO$^+ 1-0$ lines.

Our initial GBT observations (proposal 05C-043) used the Ka-band and Q-band receivers
in 2006 to cover the frequency range $34 - 48$~GHz, with bandpass calibration via 
position-switching on timescales of 1~minute. The Ka-band was found to have high 
system temperatures over $36-40$~GHz, and data from both receivers were affected by 
variable structure in the passband that could not be calibrated out. After a number of 
tests of the system stability, we re-designed the survey in proposal 07C-018, using 
the Q-band receiver to cover the frequency range $\approx 39-49$~GHz. Position-switching
on timescales of 1~minute was initially used for passband calibration, but this did 
not yield spectral baselines of the requisite quality. The second part of the survey hence 
used sub-reflector nodding, on timescales of 4.5~seconds, for passband calibration. This 
strategy was found to give smooth spectra over each 800~MHz ACS sub-band that could 
be fitted with a 5th- or 6th-order polynomial. 

The final GBT observations for the absorption survey were carried out in 2008, 
using the ACS, 3-level sampling, 2 polarizations, and two 800~MHz sub-bands, 
overlapped by 100~MHz to yield a total bandwidth of 1.5~GHz. Each 800~MHz 
sub-band was divided into 2048 channels, giving velocity resolutions of 
$4.7-5.9$~\kms\ after Hanning smoothing. For each source, we 
used seven overlapping frequency settings, each with a net bandwidth of 1.5~GHz, 
to cover the frequency range $39.6 - 49.5$~GHz. The large overheads from 
sub-reflector nodding meant that we were only able to observe 38~sources of the 
sample with this mode. Note that these were selected randomly from the full sample, 
based on whether they could be observed during the scheduled runs; as such, there was 
no bias towards brighter sources. The total integration times per source were $1-4$~minutes,
with 1.5~second records, with longer integrations on the fainter sources so as to 
obtain comparable optical depth sensitivity. The flux density scale was calibrated 
using online measurements of the system temperatures with a blinking noise diode.

The GBT data were initially analysed in the package {\sc gbtidl}, using standard 
procedures. Minor data editing was needed due to a few spectrometer failures. 
After flagging, gain and bandpass calibration, the 1.5-second records were averaged to 
produce fourteen 800~MHz spectra for each source (7 frequency settings per source, each with 
two 800~MHz ACS sub-bands), covering $\approx 39.6 - 49.5$~GHz. In all, there were 532 
such 800~MHz spectra (38~sources and 14~spectra per source), which were then analysed 
independently, outside {\sc gbtidl}. 

A polynomial of order $3-6$ was fitted to each 800~MHz spectrum (excluding edge 
channels), and subtracted out to produce a residual spectrum. Beginning with a 3rd-order 
fit, the residuals were tested for Gaussianity, using an Anderson-Darling test 
\citep[][]{anderson54}. If 
the residuals were found to be non-Gaussian ($p < 0.0001$ in the Anderson-Darling test), 
the procedure was repeated after increasing the order, upto order $6$. If even a 6th-order 
fit was found to yield non-Gaussian residuals, the spectrum was dropped from the later 
analysis. This approach was followed so as to avoid the possibility of fitting out possible 
absorption features. We verified in all cases that the fitted polynomials were smooth 
on scales $>> 100$~\kms, implying that narrow absorption lines would not be affected by 
the fitting procedure. Ninety-four spectra were excluded due to the non-Gaussianity of the 
residuals after the 6th-order polynomial fit, mostly due to ACS failures and radio 
frequency interference (RFI), especially at the upper end of the frequency range. 
The rejected 94 spectra were also inspected by eye to ensure that the non-Gaussianity did 
not arise due to the presence of spectral features.

After the above procedure, 438 spectra were retained towards 36~targets (two sources
were entirely removed). The results are summarized in Table~\ref{table:sample}, whose 
columns are (1)~the AGN name, 
(2)~the AGN redshift, for sources with known redshifts, (3)~the usable frequency 
range, after all data editing, (4)~the median RMS optical depth $\tau_{med}$ over 
the usable frequency range, and (5)~the redshift path for each source with a known 
redshift, summing over the the three CO and HCO$^+$ transitions. In passing, we note 
that two of the three sources without redshifts have barely been detected in deep 
Keck R-band images ($R = 24.5, 25.8$), while the third has $R > 26.1$
 \citep[][]{jorgenson06}; none are from the 1~Jy sample.

Fig.~\ref{fig:rms} shows the distribution of root-mean-square (RMS) optical depth noise values 
$\tau_{\rm RMS}$ across the 438 spectra. All but one of the spectra have $\tau_{\rm RMS} 
\leq 0.125$ per 6~\kms, implying that absorbers with CO/HCO$^+$ opacities $\geq 0.75$ 
would have been detected at $\geq 5\sigma$ significance. To ensure a relatively uniform 
lower limit to the sensitivity across the entire frequency (i.e. redshift) range, we chose 
$\tau_{\rm RMS} = 0.125$ as the threshold RMS optical depth noise, excluding the sole 
spectrum with a significantly higher $\tau_{\rm RMS}$ from the statistical analysis. 

The search for absorption features in the spectra with $\tau_{\rm RMS} \leq 0.125$ was 
carried out at velocity resolutions of $5-50$~\kms, smoothing the spectra to the 
resolution of interest to maximize the signal-to-noise ratio. Nine features were detected 
at $\geq 4\sigma$ significance, after integrating over all absorption channels, with four 
features detected at $\gtrsim 5\sigma$ significance.  To examine whether these 
might be due to RFI, the spectra of other sources observed on the same day were inspected
to test whether similar features were detected at the same frequency; this was 
not the case for any of the putative features. Note that, based on Gaussian noise 
statistics, we expected a total of nine features with $\geq 4\sigma$ significance 
in the spectra, but no features with $\geq 5\sigma$ significance. 

Finally, we used the VLA Q-band receivers to follow up the nine candidate absorption 
features in January~2013 (proposal 12B-408). Each target was observed for 6~minutes, using
the 8-bit samplers and two 1~GHz IF bands, with one of the 128~MHz digital sub-bands (divided 
into 512 channels) of one IF band centred on the feature in question. The second 
IF band was used to cover the frequency of a feature towards another target, to test the 
possibility of RFI. The observations of each feature were sufficiently deep to detect it 
at $\geq 10\sigma$ significance. The data were analysed in {\sc aips}, using standard 
procedures. None of the 9 candidate features seen in the GBT spectra were detected in 
the VLA spectra, indicating that they are either noise artefacts or low-level RFI. 

\begin{figure}
\centering
\includegraphics[scale=0.4]{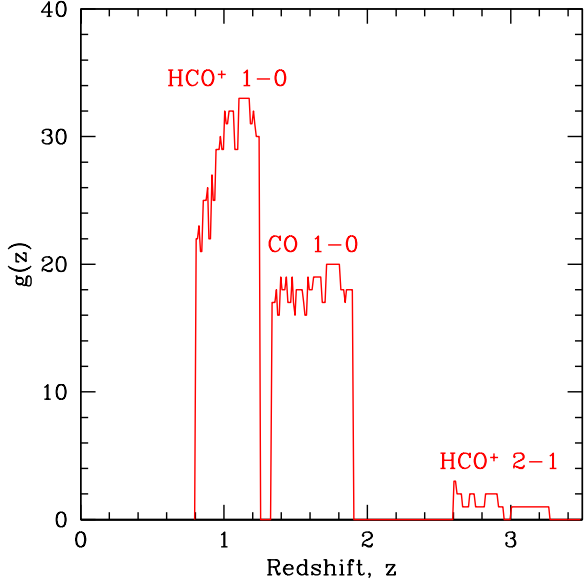}
\caption{The redshift sensitivity function of the survey $g(z)$ plotted against redshift.
The function has three separate peaks (at $z \approx 1.2$, $1.6$ and $2.8$) because 
the three transitions HCO$^+ 1-0$, CO~$1-0$ and HCO$^+ 2-1$ sample different redshift 
ranges. The maximum value of $g(z)$ is at $z \approx 1.2$.}
\label{fig:gz}
\end{figure}

\section{Discussion}
\label{sec:results}

The threshold sensitivity of $\tau_{\rm RMS} = 0.125$ implies that column densities of 
$N({\rm CO}) \approx 5 \times 10^{15}$~\cm\ and $N({\rm HCO^+}) \approx 6 \times 10^{12}$~\cm\ 
would have been detected at $5\sigma$ significance (assuming excitation temperatures of $10$~K). 
These can be converted to equivalent ${\rm H_2}$ column densities using the Galactic conversion 
factors $N({\rm CO}) \sim 3 \times 10^{-6} N({\rm H_2})$ 
\citep[valid for diffuse/translucent clouds; ][]{burgh07} and 
$N({\rm HCO^+}) \sim 2 \times 10^{-9} N({\rm H_2})$ \citep[e.g.][]{liszt00}. 
Note that the former is the median value of the ratio $N({\rm CO})/N({\rm H_2})$ obtained 
by \citet{burgh07} for diffuse/translucent clouds; dark clouds are likely to have 
lower ratios, $\sim 2-3 \times 10^{-4}$ \citep[e.g.][]{lacy94}. The use of the lower 
conversion factor is a conservative approach, as the higher value would imply that 
lower ${\rm H_2}$ column densities would have been detectable in absorption in our spectra.
We also note that the above implicitly assumes that the Galactic conversion factors,
for gas at roughly solar metallicity, may be used in high redshift galaxies. Given these 
assumptions, any molecular gas along the survey sightlines with $N({\rm H_2}) \gtrsim 
3 \times 10^{21}$~\cm\ should have been detected at $\geq 5\sigma$ significance.

Fig.~\ref{fig:gz} plots the redshift path density $g(z)$ of the present survey versus 
redshift for the 33 sources with known redshifts. The three peaks in $g(z)$ are because the 
HCO$^+ 1-0$, CO~$1-0$ and HCO$^+ 2-1$ lines sample different redshift ranges,
$0.80 - 1.21$, $1.33 - 1.91$ and $2.60-3.27$, respectively. It is clear that $g(z)$ 
peaks at $z \approx 1.2$; this is due to the selection criterion that the targets
have either $z \geq 1.2$ or no known redshift. 

The total survey redshift path is obtained by integrating the redshift path density 
over redshift, i.e.  $\Delta z = \int_0^\infty g(z) dz$. For the 33 sources with known 
redshifts, we obtain $\Delta z \approx 24$, for $\tau_{\rm RMS} \leq 0.125$ and 
$N({\rm H_2}) \gtrsim 3 \times 10^{21}$~\cm. For comparison, there have so far been 
two surveys for damped Lyman-$\alpha$ absorption (with \hi\ column densities 
$N({\rm HI}) \geq 2 \times 10^{20}$~\cm) based on complete radio-selected samples, 
the CORALS and UCSD surveys \citep[][]{ellison01,jorgenson06}. These surveys were typically
sensitive to absorbers at higher redshifts, $z \gtrsim 2$, with total redshift paths of 
$\approx 55$ and $\approx 41$, respectively, significantly larger than the redshift path 
of the present molecular absorption survey.

No molecular absorbers were detected in the present survey. Using small-number 
statistics \citep[][]{gehrels86}, the $2\sigma$ upper limit to the number of molecular 
absorbers in the survey redshift path ($\Delta z \approx 24$) is then $N < 3.7$. 
The $2\sigma$ upper limit to the redshift number density of molecular absorbers with 
$N({\rm H_2}) \geq 3 \times 10^{21}$~\cm\ at the peak survey redshift of $z \approx 1.2$ 
is thus $n(z=1.2) = (N/\Delta z) \lesssim 0.15$.

It would be interesting to compare the redshift number density of molecular absorbers
to the redshift number density of atomic absorbers at the same redshift. Unfortunately,
there have so far been no DLA surveys using radio-selected quasar samples at $z < 1.6$. In fact,
since the Lyman-$\alpha$ line is in the ultraviolet for these redshifts, the only information
that we have on the redshift number density of atomic gas is from Lyman-$\alpha$ and \hi~21cm
absorption surveys of samples selected on the basis of strong \mgtwo\ absorption 
\citep[e.g.][]{rao06,lane00b,kanekar09b}. Surveys for DLAs based on strong \mgtwo\ 
absorption have yielded $n_{\rm DLA}(z=1.219) 
\approx 0.120 \pm 0.025$ \citep[][]{rao06} for the redshift number density of DLAs, comparable 
to our upper limit on the number density of molecular absorbers. However, the biases in the 
result of \citet{rao06} are not known, because (1)~the sample of background quasars is not a 
complete radio sample, and is hence subject to dust obscuration bias, and (2)~the effect of the 
initial \mgtwo\ selection is unclear. For example, the absorption cross-section appears higher at 
${\rm log}[N($H{\sc i}$)/{\rm cm^{-2}}] = 21.6$ than at ${\rm log}[N($H{\sc i}$)/{\rm cm^{-2}}] = 21.3$ 
in the $z \approx 1$ DLAs found via \mgtwo\ selection, which is unlikely \citep[][]{prochaska09}.

One might use the upper limit on the redshift number density to derive an upper limit 
on the cosmological mass density of molecular gas at $z \approx 1.2$ \citep[following the 
arguments used in DLA surveys; e.g.][]{wolfe05}. Of course, this would require one to 
assume that the the conversion factors from CO/HCO$^+$ column densities to ${\rm H_2}$ column
densities remain unchanged with redshift, which may not be valid. In the present case, 
the lack of detections of molecular absorption implies that we have no information on 
the shape of the distribution function of the molecular gas column density; this makes 
such an exercise very speculative. Inferring the cosmological mass density of molecular gas 
as well as the relative spatial extents of the atomic and molecular gas must hence await 
similar surveys with a significantly larger total redshift path, with the VLA or the 
Atacama Large Millimeter Array. 

For 13 of the 33 targets with known redshifts, indicated by a $\dagger$ in 
Table~\ref{table:sample}, the survey also covers the AGN redshift. The lack of detected 
molecular absorption from the AGN host galaxy indicates that little molecular gas is 
present in the host galaxies along the AGN sightlines. It would be interesting to 
probe the AGN environment by examining the absorption detection fraction as a 
function of AGN type, as has been done for redshifted \hii\ absorption 
\citep[e.g.][]{vermeulen03,gupta06a}. However, this too will have to await the advent of 
larger absorption surveys.

In summary, we present the methodology for ``blind'' millimetre-wave molecular absorption
surveys in the redshifted CO/HCO$^+$ lines, and find that the $30-50$~GHz frequency range is 
optimal for such surveys, covering absorbers at redshifts $z \gtrsim 0.8$. We  have used this 
approach to carry out the first large ``blind'' survey for redshifted molecular absorption 
with the GBT, covering the frequency range $39.6-49.5$~GHz, i.e. the redshift ranges 
$0.81 - 1.27$, $1.33 - 1.91$ and $2.61 - 3.27$ in, respectively, the HCO$^+ 1-0$, CO~$1-0$ and 
HCO$^+ 2-1$ lines. Thirty-six sources were searched for absorption, 33 with known redshifts,
$z > 1.2$.  The final survey redshift path was $\Delta z \approx 24$ at an optical depth 
threshold sensitivity of $\tau_{\rm RMS} = 0.125$. We obtained no confirmed detections 
of absorption, yielding the $2\sigma$ upper limit $n(z = 1.2) < 0.15$ on the redshift 
number density of molecular absorbers with equivalent molecular hydrogen column densities 
$N({\rm H_2}) \geq 3 \times 10^{21}$~\cm. 

\begin{table*}
\caption{\label{table:sample} The 36 sources searched for millimetre-wave 
molecular absorption in the survey. The 13 sources for which the host galaxy redshift 
is covered by the absorption survey are indicated by a $\dagger$ in column~(1). 
The last column contains the total redshift path for CO/HCO$^+$ absorption towards 
sources with known redshifts, including the HCO$^+ 1-0$, CO~$1-0$ and HCO$^+ 2-1$ lines.}
\begin{center}
\begin{tabular}{|c|c|c|c|c|}
\hline
AGN &  $z$   &  Frequencies covered (GHz) & $\tau_{med}$ & $\Delta z$ \\
\hline
\multicolumn{5}{|c|}{Sources with known redshifts} \\
\hline
0202+319$\dagger$  &  1.466 & $39.6 - 49.5$                    &  $0.008$     & 0.558      \\
0223+341$\dagger$ &  2.910 & $39.6 - 49.5$                    &  $0.024$     & 1.328      \\
0234+285$\dagger$ &  1.213 & $39.6 - 49.5$                    &  $0.007$     & 0.388      \\
0248+430  &  1.310 & $39.6 - 49.5$                    &  $0.028$     & 0.447      \\
0332+078  &  1.982 & $39.6 - 42.47$, $43.15-49.5$     &  $0.056$     & 0.948      \\
0400+258  &  2.109 & $39.6 - 49.5$                    &  $0.013$     & 1.024      \\
0414-189$\dagger$  &  1.536 & $39.6 - 48.07$, $48.75 - 49.5$   &  $0.032$     & 0.568      \\
0434-188  &  2.702 & $39.6 - 42.47$, $43.15 - 43.87$  &  $0.060$     & 0.493      \\
0451-282  &  2.559 & $39.6 - 49.5$                    &  $0.024$     & 1.024      \\
0511-220  &  1.296 & $39.6 - 45.97$                   &  $0.039$     & 0.309      \\
0528+134  &  2.060 & $39.6 - 46.67$, $47.35 - 48.77$  &  $0.023$     & 0.901      \\
0528-250  &  2.813 & $39.6 - 44.57$                   &  $0.093$     & 0.569      \\
0537-286  &  3.104 & $39.6 - 42.57$                   &  $0.032$     & 0.343      \\
0602+673  &  1.970 & $39.6 - 46.67$, $47.35 - 49.5$   &  $0.042$     & 0.961      \\
0742+100$\dagger$  &  2.624 & $39.6 - 49.5$                    &  $0.031$     & 1.042      \\
0805-077$\dagger$  &  1.837 & $39.6 - 49.5$                    &  $0.037$     & 0.925      \\
0834-201$\dagger$  &  2.752 & $39.6 - 48.07$, $48.75 - 49.5$   &  $0.026$     & 1.059      \\
0839+187  &  1.270 & $39.6 - 46.67$, $47.35 - 48.07$  &  $0.055$     & 0.344      \\
0858-279  &  2.152 & $39.6 - 44.57$                   &  $0.065$     & 0.569      \\
0859-140  &  1.333 & $40.37 - 42.47$, $41.75 - 45.97$ &  $0.066$     & 0.217      \\
0919-260  &  2.300 & $39.6 - 45.97$, $46.65 - 47.37$  &  $0.039$     & 0.776      \\
1032-199  &  2.198 & $39.6 - 49.5$                    &  $0.033$     & 1.024      \\
1039+811  &  1.260 & $39.6 - 46.67$, $47.35 - 49.5$   &  $0.032$     & 0.397      \\
1150+812$\dagger$  &  1.250 & $39.6 - 49.5$                    &  $0.033$     & 0.424      \\
1354-152$\dagger$  &  1.890 & $39.6 - 45.97$, $46.65 - 48.07$  &  $0.044$     & 0.792      \\
1406-076$\dagger$  &  1.494 & $39.6 - 49.5$                    &  $0.035$     & 0.586      \\
1435+638  &  2.068 & $39.6 - 40.37$, $41.05 - 49.5$   &  $0.021$     & 0.940      \\
2126-158$\dagger$  &  3.268 & $39.6 - 44.57$, $45.25 - 46.67$  &  $0.066$     & 1.369      \\
2131-021  &  1.285 & $39.6 - 49.5$                    &  $0.009$     & 0.447      \\
2134+004  &  1.932 & $39.6 - 49.5$                    &  $0.006$     & 1.024      \\
2223-052$\dagger$  &  1.404 & $39.6 - 49.5$                    &  $0.006$     & 0.496      \\
2227-088$\dagger$  &  1.560 & $39.6 - 49.5$                    &  $0.006$     & 0.650      \\
2351+456  &  1.992 & $39.6 - 49.5$                    &  $0.025$     & 1.024      \\
\hline
\multicolumn{5}{|c|}{Sources without known redshifts} \\
\hline
0102+480  &   $-$  & $39.6 - 49.5$		      &  $0.036$     & $-$       \\
0633+596  &   $-$  & $39.6 - 49.5$		      &  $0.029$     & $-$       \\
0718+792  &   $-$  & $39.6 - 49.5$		      &  $0.033$     & $-$       \\
% 0939+140  &   $-$  & $39.6 - 49.5$		      &  $0.000$     & $-$       \\
% 1205+540  &   $-$  & $39.6 - 48.07$		      &  $0.000$     & $-$       \\
\hline
\end{tabular}
\end{center}
\end{table*}

\acknowledgments
We thank Carl Bignell, Bob Garwood and Ron Maddalena for much help with the GBT observations 
and data analysis. NK acknowledges support from the Department of Science and Technology, 
via a Ramanujan Fellowship, AG from the NCRA Visiting Students' Research Programme, 
KWW from the NRAO Graduate Summer Student Research Assistantship program, CC and 
NK from the Max-Planck Society and the Alexander von Humboldt Foundation, and JTS 
from NSF grant AST-0707480 and an NRAO travel grant. The National Radio Astronomy Observatory 
is operated by Associated Universities, Inc, under cooperative agreement with the 
National Science Foundation.  
\bibliographystyle{apj}
% \bibliography{ms}

\end{document}